\documentclass[pss]{wiley2sp} 
\usepackage{amsmath,amsfonts,amssymb}
\usepackage{txfonts}

\usepackage{bm}              
\usepackage{w-greek}         
\usepackage{units}

\usepackage[bookmarks=true,letterpaper=true,colorlinks=true,urlcolor=blue,linkcolor=blue,citecolor=blue]{hyperref}
\definecolor{Tan}{rgb}{0.737255,0.56078431,0.56078431}
\renewcommand{\copyrightmark}{}
\journalname{Phys. Status Solidi B / \textbf{DOI} 10.1012/pssb.201451171}

\begin{document}

\title{PFO-BPy Solubilizers for SWNTs: Modelling of Polymers from Oligomers}

\titlerunning{Modelling of Polymers from Oligomers}

\author{%
  Livia N. Glanzmann, 
  Duncan J. Mowbray
, and
  Angel Rubio
}

\authorrunning{L. N. Glanzmann et al.}

\mail{e-mail
  \textsf{livia.glanzmann@gmail.com}, Phone:
  +34 943 01 8392, Fax: +34 943 01 8302}

\institute{%
Nano-bio Spectroscopy Group and ETSF Scientific Development Centre, Departamento de F{\'{\i}}sica de Materiales, Centro de F{\'{\i}}sica de Materiales CSIC-UPV/EHU-MPC and DIPC, Universidad del Pa{\'{\i}}s Vasco UPV/EHU, Av.~Tolosa 72, E-20018 San Sebasti{\'{a}}n, Spain
}



\abstract{%
%
%
%
Due to their exeptional physical properties, single walled carbon nanotubes (SWNTs) embedded in organic polymers (polymer-SWNT hybrid systems) are promising materials for organic photovoltaic (OPV) devices. Already at the SWNT sorting and debundling step, polymers such as the copolymer of 9,9-dioctylfluorenyl-2,7-diyl and bipyridine (PFO-BPy) are used as solubilizers. However, to model polymer-SWNT hybrid systems, we must first determine the smallest oligomer needed to sufficiently describe the electronic and optical absorption properties of the polymer. To do so, we use time dependent density functional theory (TDDFT) to model the PFO-BPy polymer using the monomers, dimers and trimers of the PFO-BPy and Py-PFO-Py building blocks, which are also compared to the infinitely long polymer.  We find the Py-PFO-Py monomer, with shortened side chains, already describes the PFO-BPy polymer within the expected accuracies of TDDFT. 
}

\maketitle   

\section{Introduction}

\begin{figure}
\centering
 \includegraphics*[width=0.98\linewidth]{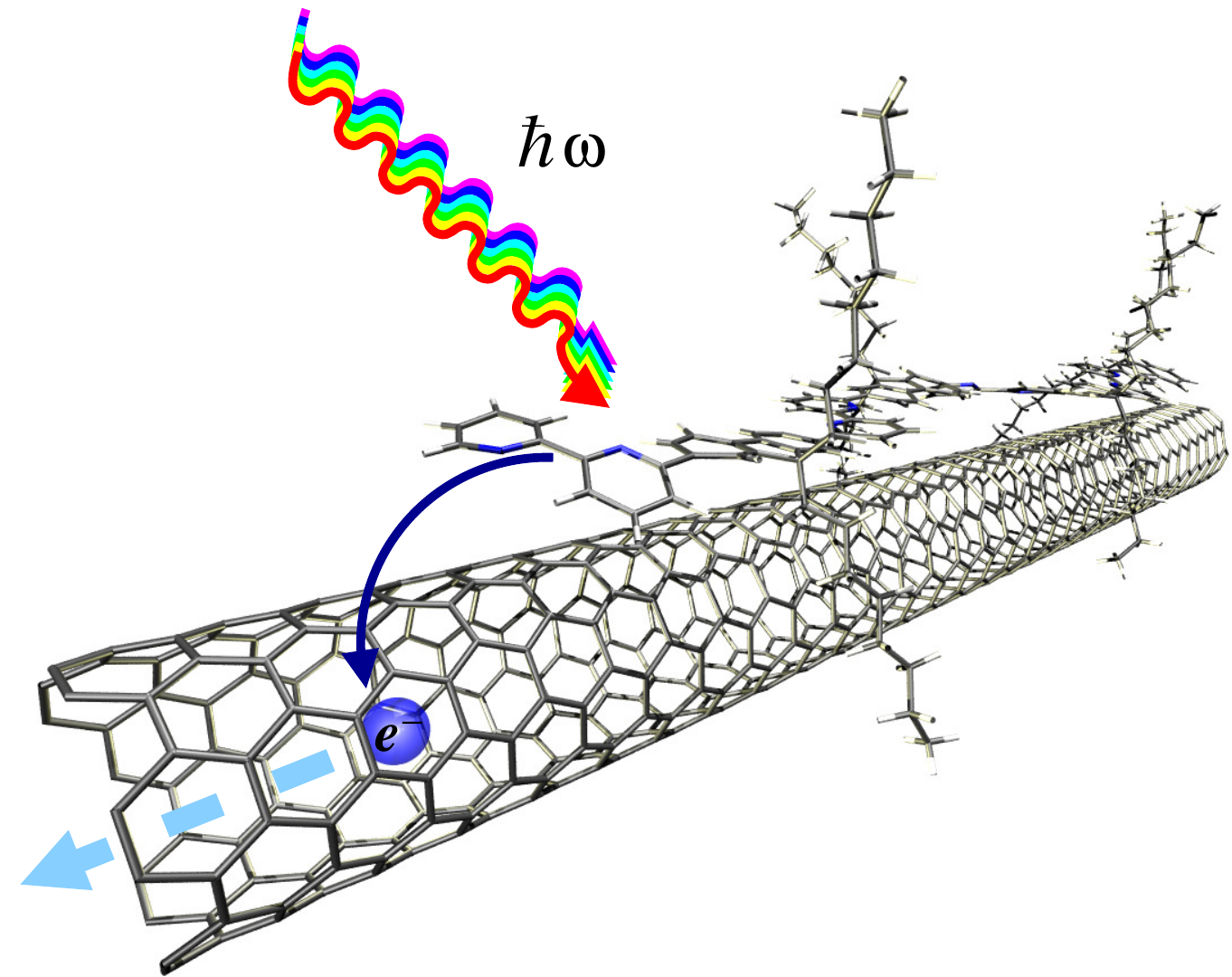}
\caption{%
Cartoon of a polymer-SWNT hybrid system as the photoactive layer in a bulk hetero junction (BHJ). The polymer absorbs light $\hbar\omega$ and transfers an electron $e^-$ to the SWNT.
}\label{Fig1}
\end{figure}

Single walled carbon nanotubes (SWNTs) \cite{Iijima,Harris,Dresselhaus} are promising acceptor materials for organic photovoltaics (OPVs) \cite{Kymakis02,Kymakis03,HertelPumpProbeSWNT+PFO-BPy}. The high surface to volume ratio of conjugated carbons in the tube structure facilitates the exciton  separation \cite{exciton} and increases the charge-carrier mobility \cite{Ferguson}. In carbon nanotube transistors, the intrinsic mobility at room temperature was found to exceed those for all known semiconductors ($>$100,000 cm$^2$/Vs) \cite{mobility}. To increase the long-lived carrier population and reduce the number of charge carrier recombination centers, enriched semiconducting SWNT solutions not containing metallic SWNTs are favoured \cite{Holt}. Addressing specific single tubes with the right band gaps allows one to tune the donor-acceptor (polymer-SWNT) interface, forming the active layer in bulk hetero junctions (BHJs) shown as a cartoon in Fig.~\ref{Fig1} \cite{Kymakis02,Kymakis03,BHJ2,Bindl}. Here, the polymer is shown absorbing light and transferring an excited electron to the SWNT.

\begin{figure}[b]
\centering
\sidecaption
 \includegraphics*[width=1.0\linewidth]{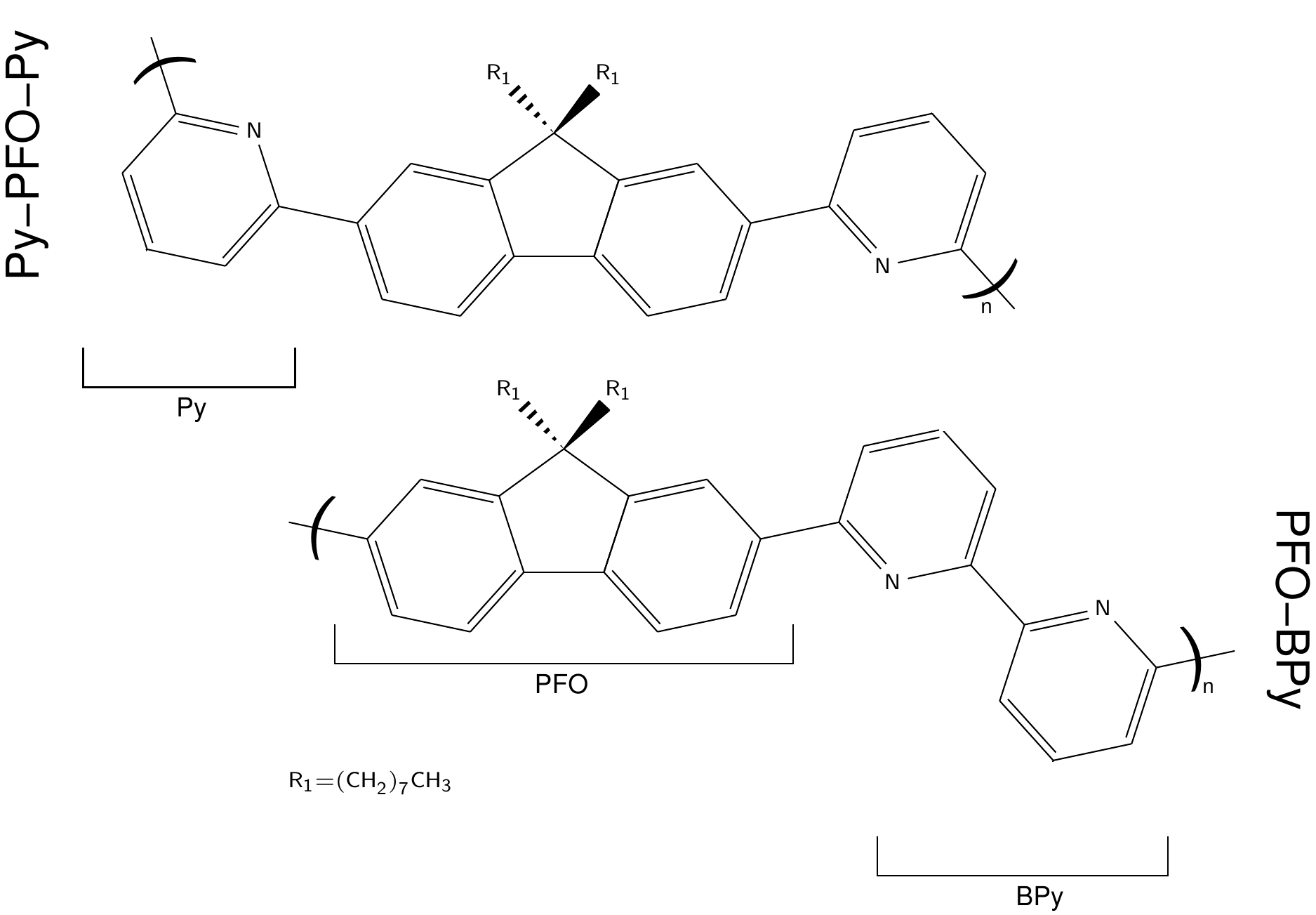}
\caption{%
Chemical structure of a pyridine-9,9-dioctylfluorenyl-2,7-diyl-pyridine (Py-PFO-Py) unit (upper panel) and the PFO-BPy unit (lower panel) of the PFO-BPy polymer.
}\label{Fig2}
\end{figure}

Thus, it is indispensable to find an efficient path to untie, sort and isolate tubes based on their diameters/chiralities. During the last ten years various approaches towards the isolation of single species SWNTs were achieved \cite{review}, such as density gradient ultracentrifugation \cite{Arnold}, current dielectrophoresis \cite{Krupke}, or the usage of solubilizers \cite{Nak1,Nak2,Nak3,Chen,Lee}. In 2011 Ozawa \emph{et al.} were able to produce a 97\% enriched dispersion of (6,5)-SWNTs using a copolymer of 9,9-dioctylfluorenyl-2,7-diyl and bipyridine (PFO-BPy) as a solubilizer \cite{Ozawa}. The chemical structure of the PFO-BPy polymer, and its two equivalent monomer units (Py-PFO-Py and PFO-BPy) are shown in Fig.~\ref{Fig2}. Sorting small band gap nanotubes with photoactive polymers would be a straightforward approach to the production of polymer-SWNT thin films for OPVs.

Still, the polymer-SWNT interface is not well understood. To shed light on this subject, we want to study the interactions between selected polymers and tubes by comparing their adsorption energies. As well, we want to compare electron densities, charge distributions, and tube band structures of isolated and hybrid systems, e.g. before and after adsorption. These studies can help guide the optimization of the polymer--SWNT level alignment for optoelectronic devices. However, to be able to perform computational studies on such hybrid systems, we must first find an oligomer with the minimal size of a polymer building block that reproduces the properties of interest of the full polymer. Only in so doing can we hope to reduce the computational costs sufficiently to subsequently model complex hybrid systems. 

\begin{figure}[!t]
\centering
\includegraphics*[width=0.8\linewidth]{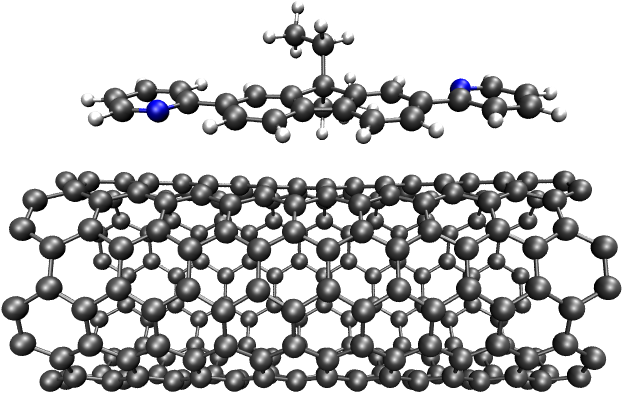}
\caption{Schematic of a Py-PFO-Py monomer adsorbed on a SWNT.}\label{Fig3}
\end{figure}

In this paper we show that the electronic properties, such as optical absorption and electon hole densities of the polymer, are already well described by the monomer unit of Py-PFO-Py, shown schematically in Fig.~\ref{Fig2}. This is accomplished by performing time-dependent density functional theory (TDDFT) calculations of the optical absorption spectra for isolated monomers, dimers, and trimers of Py-PFO-Py and PFO-BPy, and the periodically repeated polymer. Further, we compare the calculated spectra for diethyl (R$_1$ = CH$_2$CH$_3$) and dioctyl (R$_1$ = (CH$_2$)$_7$CH$_3$) side chains. These results suggest one may accurately model a complex polymer-SWNT hybrid heterojunction using the simplified monomer-SWNT hybrid system including Py-PFO-Py depicted in Fig.~\ref{Fig3}. 

\section{Methodology}\label{Methodology}

All density functional theory (DFT) calculations were performed with the real-space projector augmented wavefunction (PAW) method code \textsc{gpaw} \cite{GPAW}. We used a grid spacing of 0.2~\AA\ for representing the density and the wave functions and the PBE exchange correlation (xc)-functional \cite{PBE}.
Structural minimization was performed within the Atomic Simulation Environment (ASE) \cite{ASE},  until a maximum force below 0.05~eV/\AA\ was obtained. 

Non-periodic boundary conditions were applied for the isolated polymer units, employing more than 5~\AA\ of vacuum to the cell boundary, where both the electronic density and wavefunctions are set to zero. To model an infinitely long polymer, we repeated one Py-PFO-Py unit periodically along the polymer's length, and obtained a converged separation between repeated Py-PFO-Py units of $L = 14.866 $~\AA. 

The optical absorption spectra were calculated from the transition dipole matrix elements, which were obtained with linear response TDDFT \cite{tddft}. For these calculations we increased the number of unoccupied bands included $n_{unocc}$ to twice the number of occupied levels $n_{occ}$.  We find this resulted in a converged optical spectra up to 6~eV.  

Here, the two-point excitonic wavefunction $\Psi_n(\textbf{r}_e,\textbf{r}_h)$ for the $n^\mathrm{th}$ electron-hole pair is expressed in terms of the occupied, $\phi_i(\textbf{r}_h)$, and unoccupied, $\phi_j(\textbf{r}_e)$ Kohn-Sham (KS) wavefunctions as
\begin{equation}
\Psi_n(\textbf{r}_e,\textbf{r}_h) = \sum_{i=0}^{n_{occ}} \sum_{j=n_{occ}+1}^{n_{unocc}+n_{occ}} f_{nij} \phi_i(\textbf{r}_h) \phi_j(\textbf{r}_e),
\end{equation}
where $f_{nij}$ is the contribution of the $i\rightarrow j$ transition to the $n^{\mathrm{th}}$ excitation, so that $\sum f_{nij}^2 = 1$,  and $\textbf{r}_e$ and $\textbf{r}_h$ are the positions of the electon and hole, respectively.

The spatial distribution of a particular excitonic wavefunction $\Psi(\textbf{r}_e,\textbf{r}_h)$ may be quantified in terms of the average density of the electron $\rho_{e}(\textbf{r}_e)$ and the average density of the hole $\rho_{h}(\textbf{r}_h)$.  The electron/hole average density is obtained by integrating the density of the excitonic wavefunction with respect to the hole/electron's position, i.e.
\begin{eqnarray}
\rho_{e}(\textbf{r}_e) &=& \int\!\! d\textbf{r}_h \Psi_n(\textbf{r}_e,\textbf{r}_h)\Psi_n^*(\textbf{r}_e,\textbf{r}_h)\nonumber\\
&=& \int\!\! d\textbf{r}_h \sum_{i,i'} \sum_{j,j'} f_{ij}f_{i'j'} \phi_i(\textbf{r}_h) \phi_j(\textbf{r}_e) \phi_{i'}^*(\textbf{r}_h) \phi_{j'}^*(\textbf{r}_e)\nonumber\\
&=& \sum_{i=0}^{n_{occ}} \sum_{j=n_{occ}+1}^{n_{unocc}+n_{occ}} f_{ij}^2 \phi_j(\textbf{r}_e) \phi_{j}^*(\textbf{r}_e),\\
\rho_{h}(\textbf{r}_h) &=& \sum_{i=0}^{n_{occ}} \sum_{j=n_{occ}+1}^{n_{unocc}+n_{occ}} f_{ij}^2 \phi_i(\textbf{r}_h) \phi_{i}^*(\textbf{r}_h).
\end{eqnarray}
To compute the average electron/hole densities, we include transitions ranked by their weight $f_{ij}$ until $\sum f_{ij}^2 > 0.95$.

To compare directly with the full polymer, we have also performed calculations for the periodically repeated monomer unit. In this case, the optical absorption has been calculated from the imaginary part of the macroscopic dielectric response function $\varepsilon_m(\omega)$ obtained from linear response time dependent density functional theory within the random phase approximation (TDDFT-RPA), as recently implemented within \textsc{gpaw} \cite{response1,response2}.  The macroscopic dielectric function is calculated including local field effects by solving Dyson's equation in terms of the non-interacting density-density response function \(\chi_{\mathbf{G}\mathbf{G}'}^0(\omega)\)
\begin{equation}
\varepsilon_m(\omega) = 1\left/\left.\left[\delta_{\mathbf{G}\mathbf{G}'}+v_{\mathbf{G}\mathbf{G}'}(\mathbf{q})\chi_{\mathbf{G}\mathbf{G}'}^0(\omega)\right]^{-1}\right|_{\mathbf{G}=\mathbf{G}'=\mathbf{q}=0}\right.,
\end{equation}
where $\delta_{\mathbf{G}\mathbf{G}'}$ is the Kronecker delta, $v_{\mathbf{G}\mathbf{G}'}(\mathbf{q})$ is the Coulomb kernel, and the non-interacting density-density response function is given by
\begin{eqnarray}
\chi_{00}^0(\omega) &=& \sum_{i,j}\frac{f_i-f_{j}}{\Omega(\omega+\varepsilon_i-\varepsilon_{j}+i\gamma)}\left|\frac{\mathbf{q}\cdot\langle\phi_i|\nabla|\phi_{j}\rangle}{\varepsilon_{j}-\varepsilon_i}\right|^2,\\
\chi_{\mathbf{GG}'}^0(\omega) &=& \sum_{i,j}\frac{f_i-f_{j}}{\Omega}\frac{\langle\phi_i|e^{-i\mathbf{G}\cdot\mathbf{r}}|\phi_{j}\rangle\langle\phi_{j}|e^{i\mathbf{G}'\cdot\mathbf{r}}|\phi_{i}\rangle}{\omega+\varepsilon_{i}-\varepsilon_{j}+i\gamma},
\end{eqnarray}
where $f_i$ is the Fermi-Dirac occupation and $\varepsilon_i$ is the eigenenergy of the $i^{\mathrm{th}}$ KS wavefunction $\phi_i$, $\mathbf{G}$ and $\mathbf{G}'$ are the reciprocal lattice vectors, $\Omega$ is the unit cell volume, and $\gamma \approx 0.1$~eV.
For the calculation of the Coulomb kernel, we employ the translationally invariant Coulomb kernel which is truncated perpendicular to the unit cell \cite{CoulombCutoff},
\begin{eqnarray}
v_{\mathbf{G}\mathbf{G}'}^{\mathrm{1D}}(\mathbf{q}) &=& v_{\mathbf{G}\mathbf{G}'}^{\mathrm{3D}}(\mathbf{q})\left[1+ |\mathbf{G}_\perp|R J_1(|\mathbf{G}_\perp|R) K_0(|\mathbf{q}+\mathbf{G}_\||R)\right.\nonumber\\&&\left.\qquad\quad\ - |\mathbf{q}+\mathbf{G}_\||R J_0(|\mathbf{G}_\perp |R) K_1(|\mathbf{q}+\mathbf{G}_\|| R)\right],
\end{eqnarray}
which includes the fully periodic 3D kernel,
\begin{equation}
v_{\mathbf{G}\mathbf{G}'}^{\mathrm{3D}}(\mathbf{q}) = \frac{4\pi}{|\mathbf{q}+\mathbf{G}|}\delta_{\mathbf{G}\mathbf{G}'},
\end{equation}
where $J_0$ and $J_1$ are Bessel's functions of the first kind, $K_0$ and $K_1$ are modified Bessel's functions of the second kind, $\mathbf{G}_\perp$ and $\mathbf{G}_\|$ are the perpendicular and parallel components of $\mathbf{G}$ relative to the periodic direction, $\mathbf{q}$ is the momentum transfer ($\mathbf{q}\rightarrow 0^+$), and $R$ is the Coulomb cutoff.  To ensure the 1D Coulomb kernel retains all interactions within the unit cell, while excluding spurious interactions between periodic images, we have doubled the unit cell dimensions in the non-periodic directions, ``padding'' the KS wavefunctions with zeros, and employed a Coulomb cutoff $R$ equal to the original unit cell dimensions in the non-periodic directions \cite{MowbrayZeroPadding}.  



\section{Results \& Discussion}\label{Results}

\begin{figure}[!t]
\centering
\sidecaption
\includegraphics*[width=0.8\linewidth]{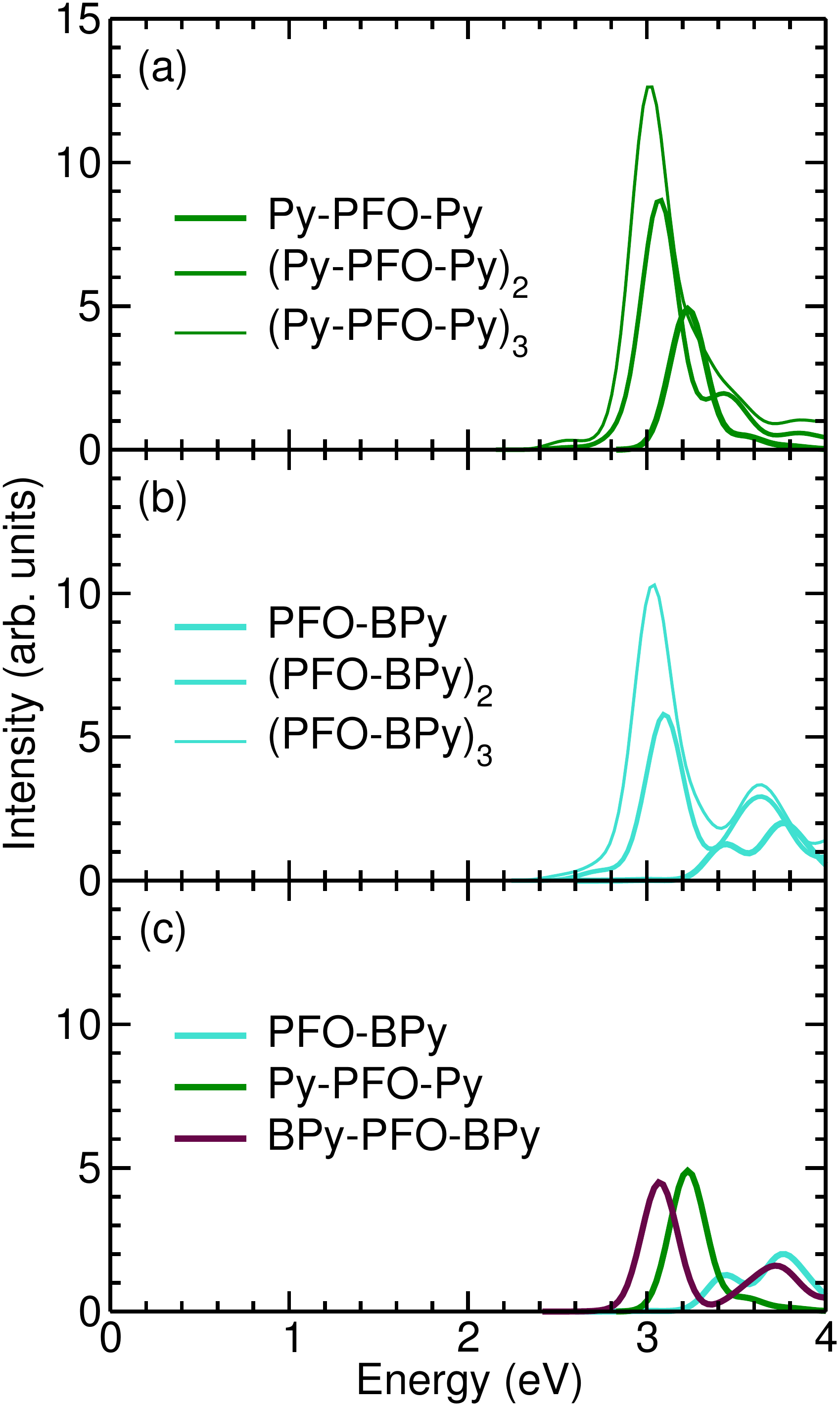}
\caption{TDDFT absorption spectra of the (a) monomer Py-PFO-Py, dimer (Py-PFO-Py)$_2$, and trimer (Py-PFO-Py)$_3$, (b) monomer (PFO-BPy), dimer (PFO-BPy)$_2$, and trimer (PFO-BPy)$_3$, and (c) monomers of PFO-BPy and Py-PFO-Py and the extended monomer BPy-PFO-BPy.}\label{Fig4}
\end{figure}

The TDDFT spectra for the monomer, dimer, and trimer of the PFO-BPy polymer are shown in Fig.~\ref{Fig4}, based on the (a) Py-PFO-Py and (b) PFO-BPy units. In both cases there is no significant redshift from the dimer to the trimer ($\sim 0.05$~eV). The absorption peak of the low energy transition band of the monomer PFO-BPy is blueshifted up to the absorption of the bipyridine part. In comparison to the peak maximum of the monomer, Py-PFO-Py is found only 0.15~eV higher in energy than its corresponding dimer. This is in contrast to PFO-BPy, which is more than 0.2~eV higher in energy.  Perhaps more importantly, the relative intensities of the two lowest energy peaks differ qualitatively between the PFO-BPy monomer and either the dimer or trimer.

In Fig.~\ref{Fig4}(c) we compare the spectra of the two monomers with the spectrum of the extended monomer BPy-PFO-BPy. The BPy-PFO-BPy monomer has on each side an additional terminating pyridine forming bipyridine. With the inclusion of these terminating pyridine units in BPy-PFO-BPy, the peak maximum of the spectrum is found to be identical to the dimer (Py-PFO-Py)$_2$. This suggests there is only a rather weak $\pi$-conjugation between subunits of the polymer.

\begin{figure}
\centering
\includegraphics[width=0.8\columnwidth]{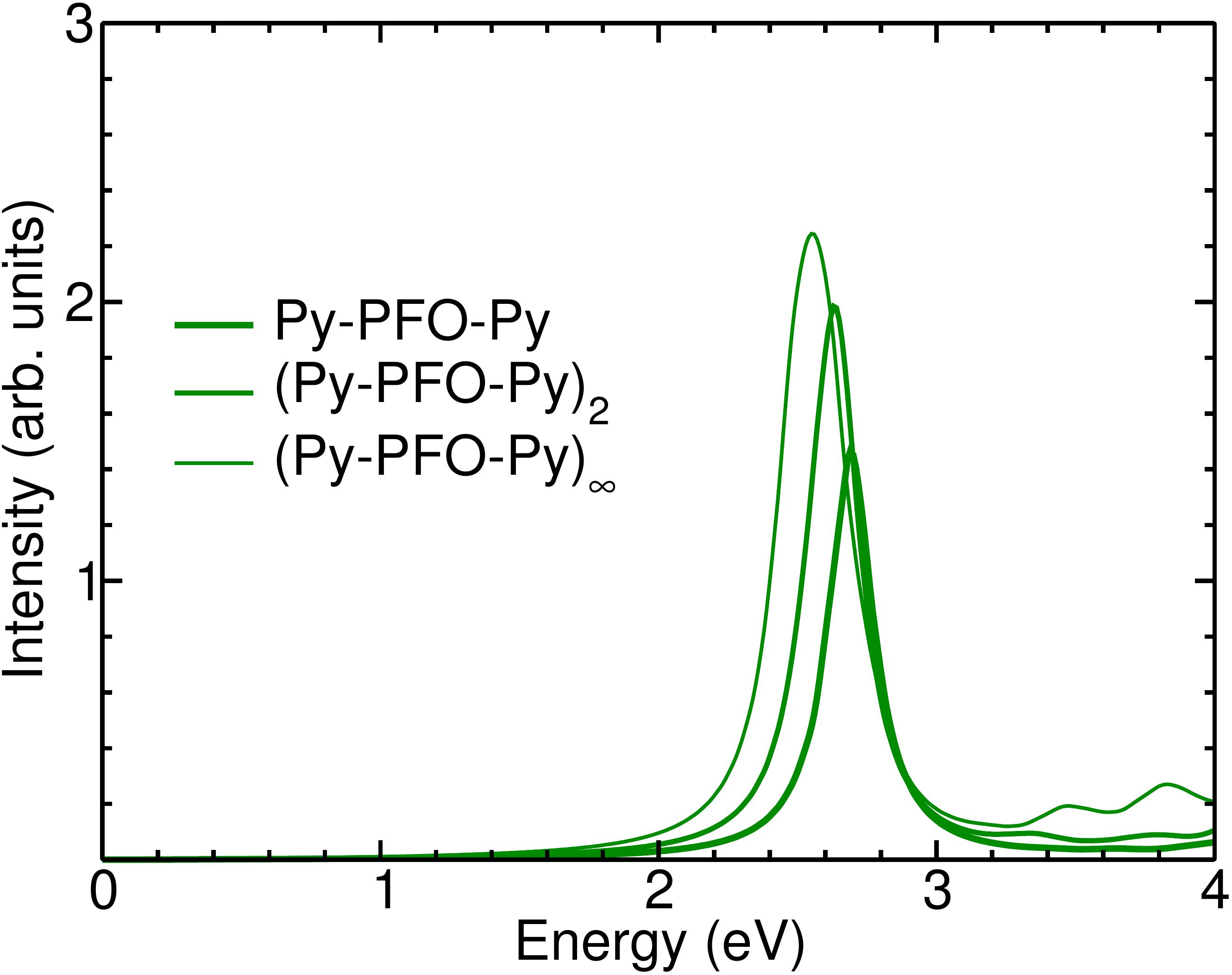}
\caption{TDDFT-RPA absorption spectra, $\Im[\varepsilon_m(\omega)]$, 
 for the monomer Py-PFO-Py, dimer (Py-PFO-Py)$_2$, and repeated polymer (Py-PFO-Py)$_\infty$.}\label{Fig5}
\end{figure}

The TDDFT-RPA absorption spectra are shown in Fig.~\ref{Fig5} for the periodically repeated polymer (PFO-BPy)$_\infty$, dimer (PFO-BPy)$_2$ and monomer (PFO-BPy). Although, all spectra are redshifted up to 0.5 eV compared to linear response TDDFT shown in Fig~\ref{Fig4}(a), we still find the oligomer spectra quickly converged to that of an infintely long polymer. Furthermore, the monomer and the dimer peak positions differ by only 0.1 eV, which is even less than with linear response TDDFT, as shown in Fig.~\ref{Fig4}(a). Moreover, the difference between the dimer and the infinte chain is around 0.08 eV. 

Within linear response TDDFT, 90\% of the brightest excitation is a HOMO-LUMO transition. Given a calculated HOMO-LUMO band gap of 2.6 eV for the monomer, TDDFT-RPA reproduces a fully HOMO-LUMO transition, whereas linear response TDDFT includes more higher lying terms.  

\begin{figure}[!t]
\sidecaption
\includegraphics*[width=0.5\linewidth]{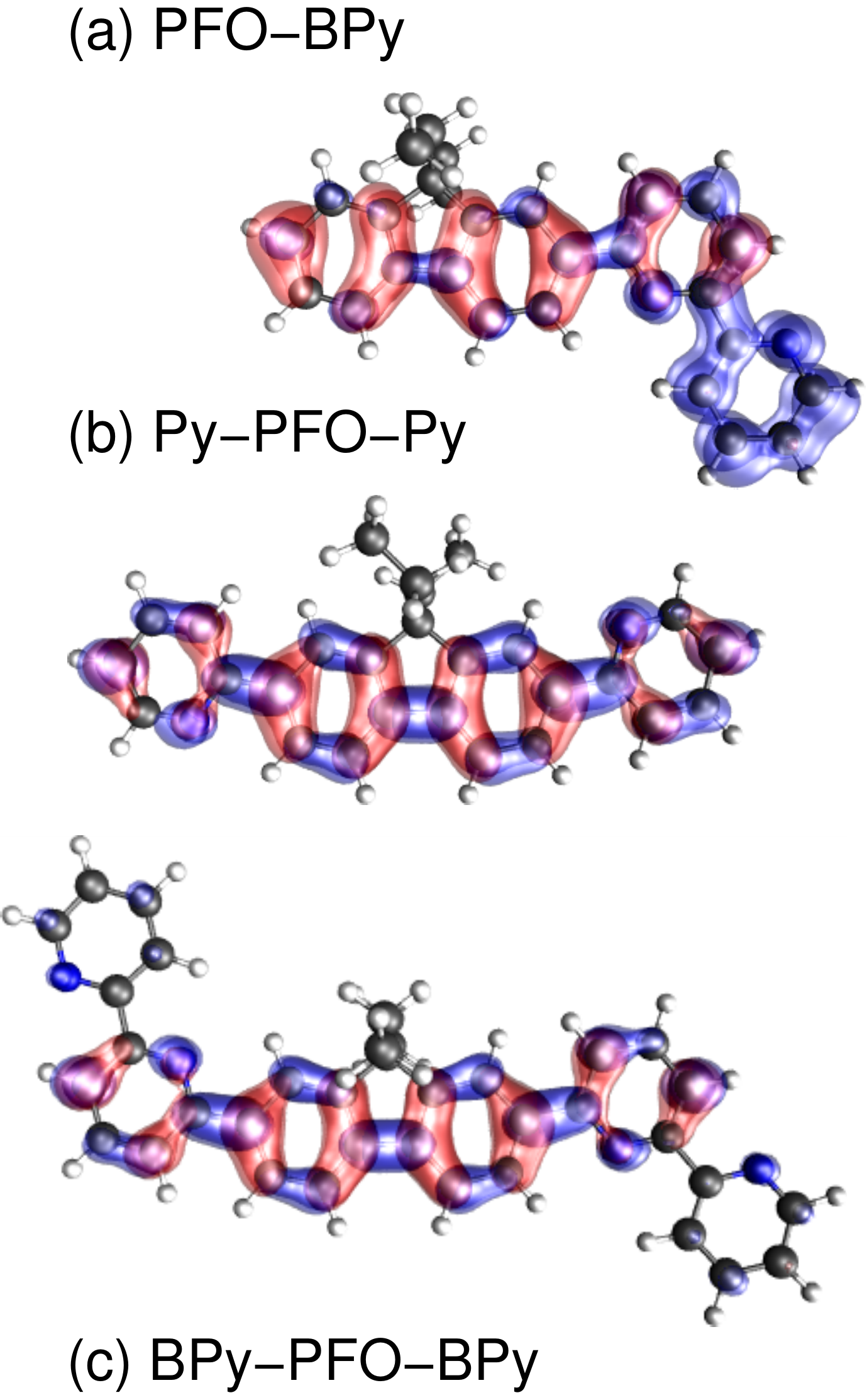}
\caption{Isosurfaces of the average electron (blue) and hole (red) densities, $\rho_e(\textbf{r}_e)$ and $\rho_h(\textbf{r}_h)$, of the exciton belonging to the most intense transition of the low energy absorbance band, i.e.\ the peak maximum, for (a) PFO-BPy, (b) Py-PFO-Py, and (c) BPy-PFO-BPy.}\label{Fig6}
\end{figure}

To understand the spatial distribution of the exciton, we next consider the averaged densities of the electron and hole within the most intense low energy peak in the spectra shown in Fig.~\ref{Fig4}(c).  
In Fig.~\ref{Fig6} the isosurfaces of the average density of the hole (red) and the average density of the electron (blue) for the most intense excitonic transition of the monomers PFO-BPy (a), Py-PFO-Py (b), and the extended monomer BPy-PFO-BPy (c) are plotted. For all the monomers, the hole is mainly located on the PFO with small weights on the neighbouring pyridines. For the monomer PFO-BPy, the density of the electron is almost only located on the bipyridine, whereas for Py-PFO-Py and BPy-PFO-BPy it is distributed over the same space as the hole densities within the antibonding $\pi$-system. More importantly, we clearly see a localization of the electron-hole pair on the central Py-PFO-Py portion of the BPy-PFO-BPy unit, which has only a minor weight of the electron on the additional terminating pyridines in BPy-PFO-BPy. This suggests that the PFO-BPy polymer lacks significant $\pi$ conjugation between Py-PFO-Py units.  

\begin{figure}[!t]
\sidecaption
\includegraphics*[width=\linewidth]{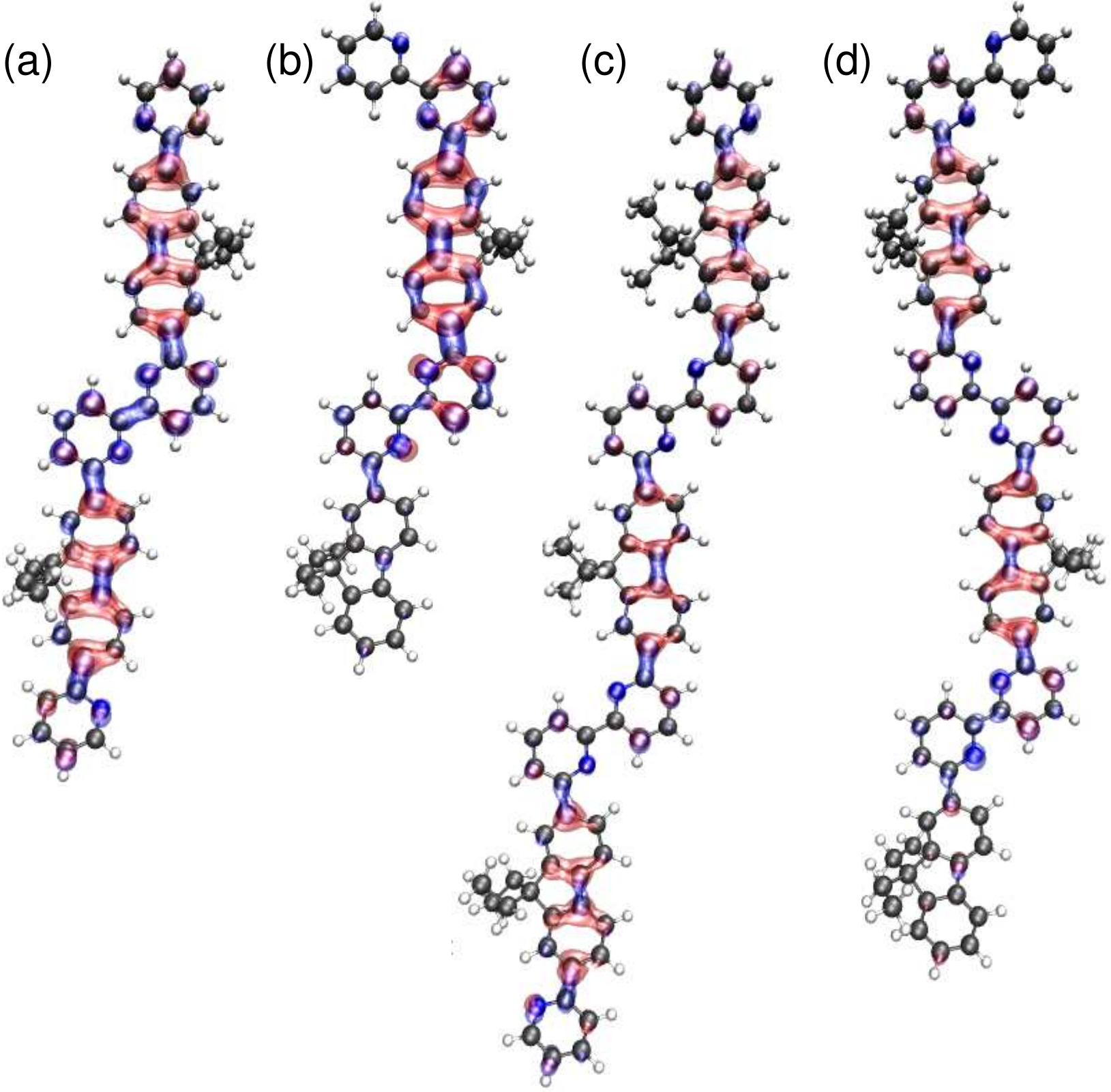}
\caption{Isosurfaces of the average electron (blue) and hole (red) densities, $\rho_e(\textbf{r}_e)$ and $\rho_h(\textbf{r}_h)$, of the exciton belonging to the most intense transition of the low energy absorbance band, i.e.\ the peak maximum, for (a) (Py-PFO-Py)$_2$, (b) (PFO-BPy)$_2$, (c) (Py-PFO-Py)$_3$, and (d) (PFO-BPy)$_3$.}\label{Fig7}
\end{figure}

The dimer (Py-PFO-Py)$_2$, and the trimer (Py-PFO-Py)$_3$ in Fig.~\ref{Fig7} show the same excitonic structure as the latter oligomers repeating with every attached unit. In the dimer, there is some weight of the electron densities on the pyridines, whereas for the trimer this is almost completely lost.
The exciton and the absorption band of the polymer seem to be composed of a sum of repeating excitons located on the Py-PFO-Py units. For the dimer (PFO-BPy)$_2$ and the trimer (PFO-BPy)$_3$, it is found that there are fewer excitons. Since for PFO-BPy units there is no stabalizing pyridine next to the PFO on one end, there is no exciton on this side contributing to that transition. Nevertheless, there is almost no blueshift of these larger PFO-BPy oligomers compared to the Py-PFO-Py oligomers, as shown in Fig.~\ref{Fig4}(a) and (b).  

\begin{figure}[!t]
\centering
\includegraphics*[width=0.8\linewidth]{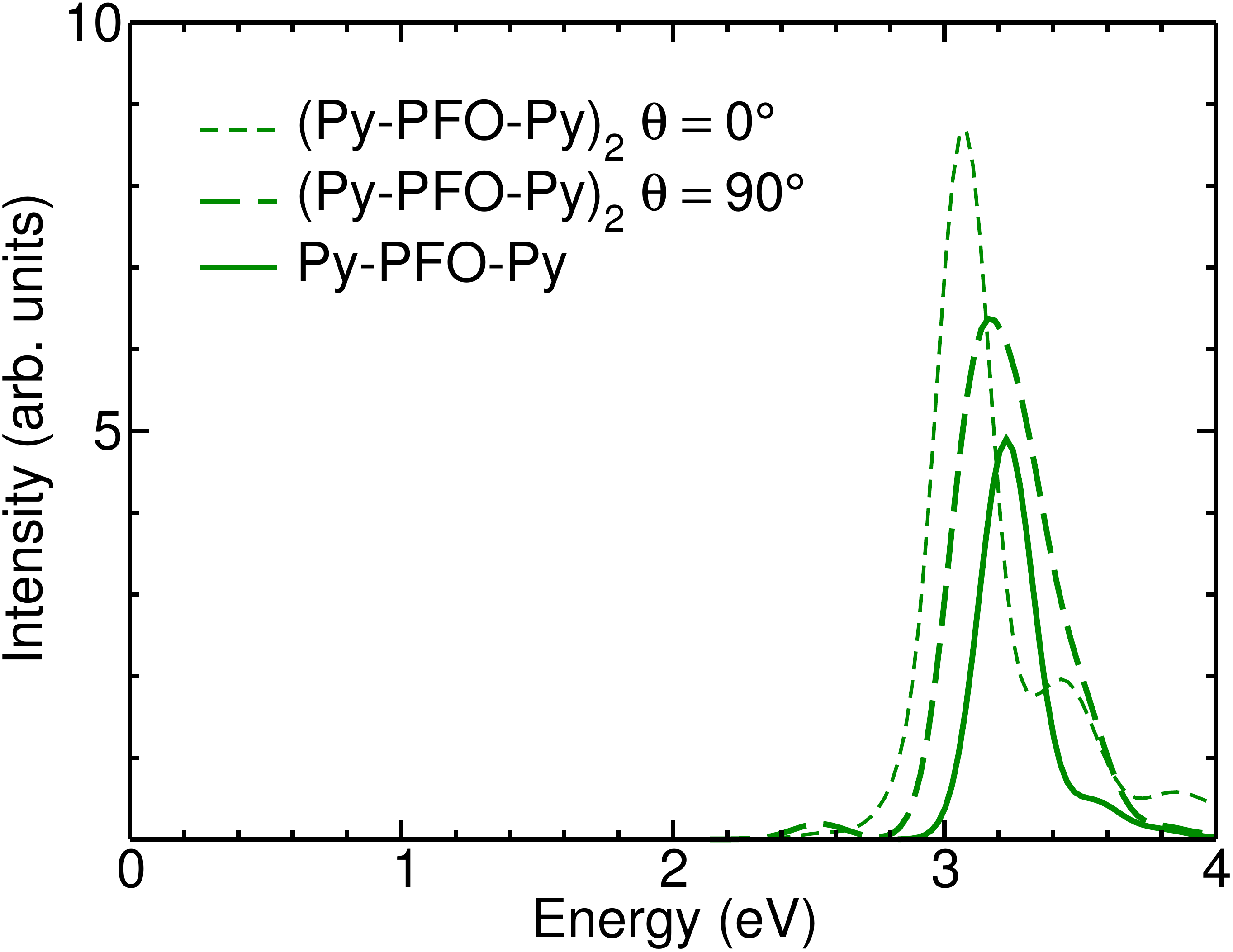}
\caption{TDDFT absorption spectra of planar (Py-PFO-PY)$_2$, i.e.\ $\theta = 0^\circ$, (Py-PFO-PY)$_2$ having the two pyridines in the bipyridine turned 90 degrees, i.e.\ $\theta = 90^\circ$, and the Py-PFO-Py monomer.}\label{Fig8}
\end{figure}

In Fig.~\ref{Fig8} we compare the absorption spectra for the dimer (Py-PFO-Py)$_2$ and monomer Py-PFO-Py spectra with that of a (Py-PFO-Py)$_2$ dimer which has been twisted to break its $\pi$-conjugation. Specifically, we rotated the Py-Py bond of the dimer (Py-PFO-Py)$_2$, twisting both monomer units 90 degrees to each other ($\theta = 90^\circ$). In this way, one Py-PFO-Py unit is perpendicular to the other. The resulting spectra of the dimer (Py-PFO-Py)$_2$ $\theta = 90^\circ$ now consists of two components. The spectrum maxima lies between the peak maxima of dimer (Py-PFO-Py)$_2$ $\theta = 90^\circ$ and the monomer Py-PFO-Py, forming a broad band which includes the monomer spectrum.  

By completely breaking the $\pi$-conjugation of the dimer, we obtain an average spectra which is much closer to that of the monomer.  This indicates that the small redshift of $\sim 0.15$~eV of the dimer (Py-PFO-Py)$_2$ is related to the small degree of $\pi$-conjugation between the monomer units seen in Fig.~\ref{Fig7}(a). This shift is also captured by using a BPy-PFO-BPy monomer model, as shown in Fig.~\ref{Fig4}(c). Here, it is related to an extension of the exciton's electron into the terminating Py rings, shown in Fig.~\ref{Fig4}(c).

\begin{figure}[!t]
\centering
\sidecaption
\includegraphics*[width=0.8\linewidth]{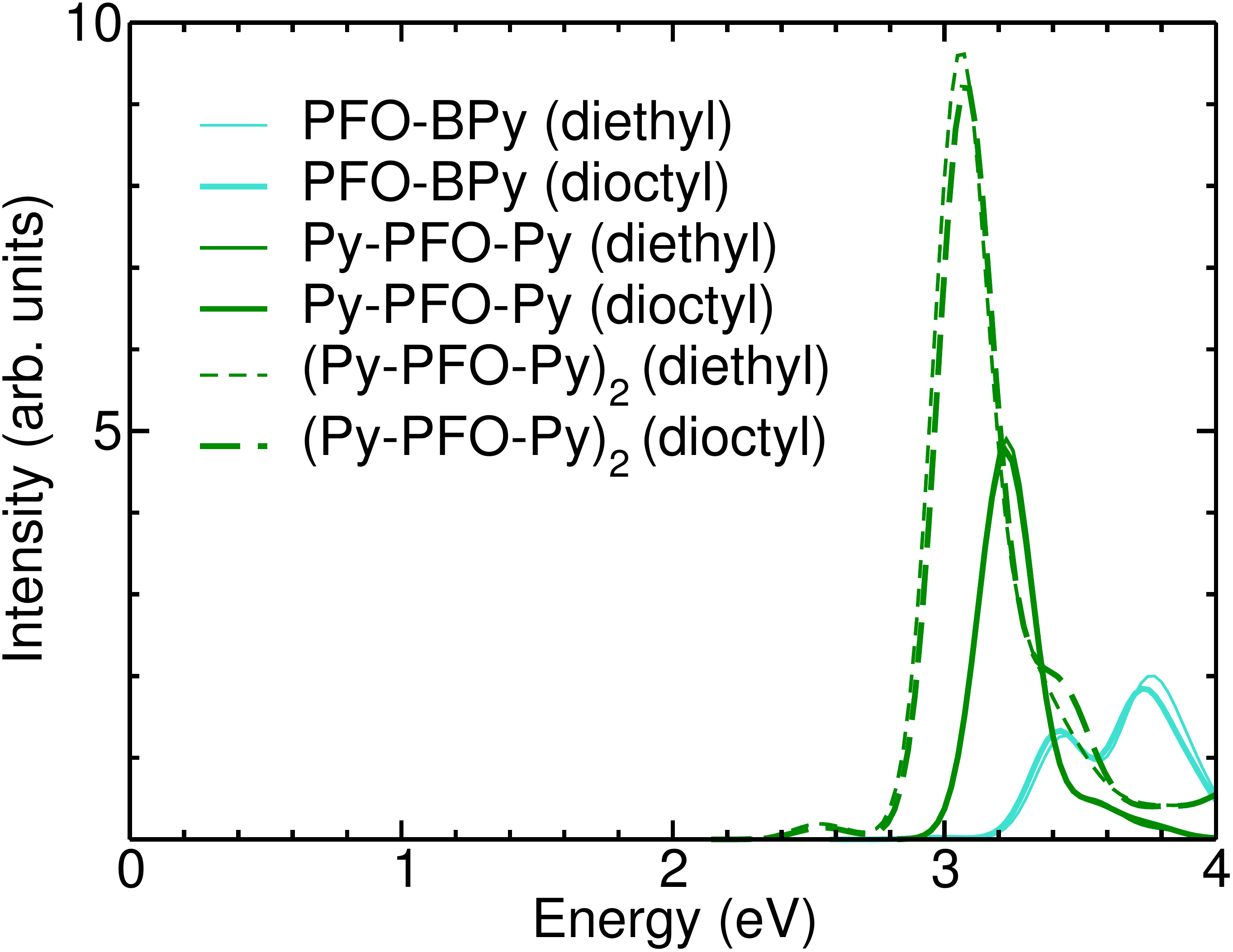}
\caption{TDDFT spectra of Py-PFO-Py (green solid line), (Py-PFO-Py)$_2$ (green dashed line), and PFO-BPy (blue line) with side chains of dioctyl (R$_1$ = CH$_2$CH$_3$, thick lines) and (R$_1$ = CH$_2$CH$_3$, thin lines) sidechains (thin line) on the PFO.}\label{Fig9}
\end{figure}

To test whether a Py-PFO-Py monomer model would include the steric influence of the side chains on the adsorption on the tube, we reduced the length of the side chain to an ethyl group. The spectra for both monomers as well as for the dimer (Py-PFO-Py)$_2$ show no peak shift in the absorption spectrum after reducing the length of the side chains, as shown in Fig.~\ref{Fig9}. These results clearly demonstrate that the length of the side chains has little influence on the optical absorption spectra of the polymer.

This is consistent with what was found by Namal \emph{et al.}\ while studying the effect of alkyl chain length on the electrochemical properties of fluorene and benzimidazole containing conjugated polymers \cite{Imge}. In this case, the first absorption band in the visible at 425 nm experiences a blueshift of 20 nm when the length of the side chains is doubled. The slight shift and the measured change of the optical band gap by only 0.08 eV is ascribed to a better solubility of the polymer and a stronger conjugation of the $\pi$--system \cite{Imge}.

\section{Conclusions}\label{Conclusions}

We have found that there is almost no redshift occurring while extending the $\pi$-system through the attachment of Py-PFO-Py oligomer units. This suggests that conjugation of the $\pi$-system is broken due to the ortho conjugation over the nitrogens of the bipyridine. As well, the electron and hole densities reveal that the exciton is only located on the Py-PFO-Py monomer unit in the polymer. On the other hand, the PFO-BPy monomer is electronically completely different from either the dimer or the trimer. By extending the PFO-BPy oligomers, the PFO-BPy system converges to the Py-PFO-Py system, albeit more slowly.  We have also confirmed that the oligomer spectra converges to that of the infintely long polymer chain using TDDFT-RPA.  We find that TDDFT-RPA underestimates the transition more than linear response TDDFT, and the TDDFT-RPA HOMO-LUMO absorption peak coincides precisely with the DFT HOMO-LUMO energy gap.

The small shift in the spectra between PFO-BPy and Py-PFO-Py systems indicates there is only a weak conjugation between the single monomer units. By breaking the conjugation through a rotation of the Py-Py bond in the dimer, the dimer only experiences a small blue shift relative to the monomer. This suggests that the shifts might as well occur due to the electron donating effect of the second pyridine considering the neighboring units as functional groups.
Additionally, we confirmed that there is no electronic influence of the octyl side chains in PFO, beyond the first ethyl group, as expected. This is consistent with previous studies of the polymer chain length's effect on the absorption properties of such polymers \cite{Imge}.

Overall, we find the Py-PFO-Py unit with shortened ethyl side chains describes the PFO-BPy polymer sufficiently well to be used when modelling electronic and optical absorption properties of a hybrid polymer-SWNT system. These results have important implications for the reliability of such simplified models for describing polymer-SWNT heterojunctions \cite{FlorianJACS}.

\begin{acknowledgement}
We thank Florian Sp\"{a}th and Tobias Hertel from the University of W\"{u}rzburg for stimulating discussions. We acknowledge funding from the European Projects DYNamo (ERC-2010-AdG-267374), CRONOS (280879-2 CRONOS CP-FP7) and POCAONTAS (FP7-PEOPLE-2012-ITN-316633); Spanish Grants (FIS2012-37549-C05-02, FIS2010-21282-C02-01, PIB2010US-00652, JCI-2010-08156); and Grupos Consolidados UPV/EHU del Gobierno Vasco (IT-319-07).

  \end{acknowledgement}

%
\bibliographystyle{pss}
\bibliography{bibliography}

\end{document}